\documentclass[conference]{IEEEtran}
\IEEEoverridecommandlockouts
\usepackage{cite}
\usepackage{amsmath,amssymb,amsfonts}
\usepackage{algorithmic}
\usepackage{graphicx}
\usepackage{textcomp}
\usepackage{xcolor}
\usepackage{booktabs}
\def\BibTeX{{\rm B\kern-.05em{\sc i\kern-.025em b}\kern-.08em
    T\kern-.1667em\lower.7ex\hbox{E}\kern-.125emX}}
    
\begin{document}

\title{Layer-wise Analysis for Quality of Multilingual Synthesized Speech}

\author{
Erica Cooper$^1$, Takuma Okamoto$^1$, Yamato Ohtani$^1$, Tomoki Toda$^{1,2}$, Hisashi Kawai$^1$
\\
\textit{$^1$National Institute of Information and Communications Technology}, \textit{$^2$Nagoya University}\\
\texttt{ecooper@nict.go.jp}
\thanks{This work was partially supported by JSPS KAKENHI Grant Number JP25K00143.}
}

\maketitle

\begin{abstract}
While supervised quality predictors for synthesized speech have demonstrated strong correlations with human ratings, their requirement for in-domain labeled training data hinders their generalization ability to new domains.  Unsupervised approaches based on pretrained self-supervised learning (SSL) based models and automatic speech recognition (ASR) models are a promising alternative; however, little is known about how these models encode information about speech quality.  Towards the goal of better understanding how different aspects of speech quality are encoded in a multilingual setting, we present a layer-wise analysis of multilingual pretrained speech models based on reference modeling. We find that features extracted from early SSL layers show correlations with human ratings of synthesized speech, and later layers of ASR models can predict quality of non-neural systems as well as intelligibility.  We also demonstrate the importance of using well-matched reference data.
\end{abstract}


\section{Introduction}
Text-to-speech (TTS) synthesis has made tremendous progress in recent years due to advancements in end-to-end modeling, speech representations, and neural vocoders \cite{wavenet,taco2,hifigan,jets}.  However, evaluation still largely relies on listening tests, especially using the Mean Opinion Score (MOS) \cite{p800} paradigm in which listeners rate speech sample quality on an Absolute Category Rating (ACR) scale.  The time-consuming and costly nature of these tests 
 has resulted in growing interest in more automated methods.  

While large-scale public MOS datasets \cite{BVCC,somos} have enabled the development of data-driven, machine-learning-based approaches that can predict MOS with high correlations to human ratings \cite{cooper2022generalization,saeki22c_interspeech},  predictors based on supervised training remain heavily task-specific and have trouble generalizing to unseen systems, languages, and domains \cite{cooper2022generalization,vmc2023}, which is especially detrimental in the case of low-resource languages for which little or no MOS-labeled data may be available.  Unsupervised approaches not requiring MOS-labeled data are a promising direction for quality prediction for speech synthesis in new domains and for low-resource languages, and large-scale pretrained multilingual speech models have been a valuable resource for a multitude of downstream tasks including quality prediction.  However, little is known about how these pretrained models encode different aspects of speech quality, which is necessary prerequisite knowledge for building the next generation of unsupervised, task-independent, and language-independent speech quality predictors.  In this paper, we take a layer-wise analysis approach to probe the layers of various pretrained models to gain a better understanding of the information they inherently contain about speech quality, which is a necessary step towards the future goal of using this information for unsupervised evaluation of synthesized speech.  

\subsection{Related Work}

We introduce some relevant past work related to automatic assessment of synthesized speech, including multilingual speech quality prediction and unsupervised speech evaluation.

\subsubsection{Multilingual synthesized speech quality prediction}

The NISQA-TTS MOS predictor \cite{mittag20_interspeech}, published in 2020, is a CNN-LSTM architecture trained on several multilingual MOS datasets including English, Chinese, German, and nine Indic languages.  This predictor demonstrated strong predictive ability even for unseen synthesis systems.  A 2022 study \cite{cooper2022generalization} built MOS predictors by fine-tuning pretrained self-supervised learning (SSL) based speech models on a large English MOS dataset, and then ran these predictors in both a zero-shot manner and with additional fine-tuning on MOS datasets in other languages to investigate their generalization ability.  Fine-tuning improved predictive ability over the zero-shot condition, but even without fine-tuning, SSL-based models demonstrated better predictive ability than non-SSL-based ones, indicating the promise of SSL models for predicting quality of synthesized speech in  unseen languages.  The SQuId predictor \cite{sellam2023squid} was an SSL model fine-tuned on proprietary MOS data from 52 different language locales.  The test set included several ``zero-shot'' locales unseen in training.  A transfer-learning effect was observed, demonstrating the benefits of training on a large variety of languages.

\subsubsection{Unsupervised evaluation of synthesized speech}

Large-scale labeled MOS datasets are few and difficult to collect for new languages and domains, and therefore several efforts have been made to move away from the task- and language-specificity of supervised approaches towards unsupervised methods.  In this paper, we use the term ``unsupervised'' to refer to approaches that are developed without using any {\em MOS-labeled} data.  Unsupervised speech quality prediction is based on the idea of a {\em reference model} learned from a distribution of natural speech samples. Quality
prediction may be achieved by measuring the distance
between the distribution of synthesized speech and that of the natural speech defined by the reference model.  

We emphasize that the use of a {\em reference model} is different from the use of {\em reference samples} in double-ended (intrusive) quality prediction methods (e.g., \cite{kubichek1993,pesq}), where the reference sample and synthesized sample must have the same lexical content, and the reference sample is considered to be the top-quality exemplar of that sentence.  These reference {\em sample} based approaches were developed for evaluating natural speech under noisy and degraded conditions for telephony applications, and are generally considered unsuitable for evaluating synthesized speech \cite{cooper2024review} because of the so-called ``one-to-many'' problem, meaning that a given sentence may be spoken in several different but acceptable ways.  In reference {\em modeling} based approaches, the natural speech samples used to create a reference model are {\em not} required to have matching lexical content to the synthesized sentences under evaluation.

In the days of hidden Markov model (HMM) based speech synthesis, Gaussian mixture models (GMMs) of distributions of acoustic features of TTS training corpora were built in one study \cite{maguer2013evaluation}, and likelihood of the reference GMMs with respect to the acoustic space of synthesized speech was proposed as an objective measure of synthesis quality.  More recently, self-supervised models including large-scale pretrained ones have been utilized for this task.  For example, SpeechLMScore \cite{speechlmscore} extracts token sequences from the third or fourth layer of a pretrained HuBERT \cite{hsu2021hubert} model followed by k-means clustering, and perplexity of the token sequences is measured with respect to a speech language model (LM) trained on natural speech.  UNIQUE \cite{yoon2024unique} takes a similar approach, using WavLM \cite{chen2022wavlm} layers 6 and 24 for feature extraction and a distribution model of k-means tokens. VQScore \cite{fuVQscore}
trains a VQ-VAE on natural clean English speech and then reconstruction error is used as a measure of quality, demonstrating improved prediction over SpeechLMScore with lower data requirements.  TTSDS \cite{minixhofer2024ttsds} uses an ensemble of reference models to measure different aspects of speech quality such as prosody, speaker identity, and intelligibility.  In one study investigating the zero-shot predictive ability of pretrained SSL models \cite{aditya}, the entropy of the token distributions were shown to have moderate correlations with MOS for datasets in several languages.  

We also note the related literature on Fréchet Inception Distance (FID) and Fréchet Audio Distance (FAD).  FID was proposed in 2017 as an evaluation metric for image generation models \cite{FID} -- it compares the distribution of two datasets in a given embedding space using the 2-Wasserstein distance.  FID originally used embeddings from the last pooling layer of the Inception-V3 \cite{inceptionv3} model, and the later-proposed FAD \cite{FAD} used embeddings from VGGish \cite{VGGish} to evaluate music enhancement algorithms.  FAD was later adapted to a TTS-specific evaluation in \cite{GANTTS}, which used DeepSpeech2 \cite{deepspeech2}, an ASR model for English, to obtain embeddings which were also extracted from the last layer.  

To the best of our knowledge, the final layer of a pretrained model is typically used for FID/FAD, but since many of these models were pretrained for an objective that is related to but not exactly the same as what is being evaluated, we consider it necessary to explore all model layers.  

\subsection{Contributions}

To the best of our knowledge, this is the first layer-wise analysis of pretrained models specifically for speech quality of synthesized speech.  We focus on a multilingual setting in order to identify which layers contain signals that correlate with human ratings of speech quality regardless of the language.

Recent unsupervised approaches for MOS prediction are  largely English-centric, using English-pretrained SSL models and conducting few multilingual evaluations (with the exceptions of VQScore, which, although it was trained on English data only, was also evaluated on Chinese-language data with good results, as well as the study investigating entropy as a predictor \cite{aditya} which also conducted evaluations on several languages).  While multilingual pretrained SSL models are available \cite{mhubert,xlsr53} and could be used to develop predictors for different languages using the unsupervised approaches cited above, these existing approaches still have heavy additional data requirements, with SpeechLMScore requiring an additional 20 thousand hours of English speech to train the speech LM and VQScore requiring 460 hours of clean English speech -- even without MOS labels, these data requirements are prohibitive for low-resource languages.  For fair and inclusive access to the benefits of speech quality prediction, it is important to consider so-called ``zero-shot'' measures of speech quality for which no additional training is required, with or without MOS labels.  

Furthermore, the model layers to use for feature extraction have typically been chosen based on prior work about topics other than speech quality prediction. More work needs to be done to identify the best pretrained models and model layers for assessing speech quality in a training-free, language-independent manner.  To this end, we conduct layer-wise analysis of several pretrained multilingual speech models to identify which models and layers correlate best with listener ratings.  We compare the layer-wise correlations with those obtained from several existing representative predictors on a variety of multilingual speech synthesis datasets in order to quantify the predictive capabilities and gaps for different languages, and we analyze the predictive ability with respect to the choice of reference data as well.

\section{Method}

Considering the use case of MOS predictors as a useful tool for speech synthesis research and development, our approach uses speech synthesis training corpora as reference data, following \cite{maguer2013evaluation}.  This is a realistic scenario since speech synthesis researchers will always have access to their own training corpus, even if no additional data in the language of interest is available.

We introduce a layer-wise analysis method for measuring similarity of extracted feature distributions of synthesized speech compared to feature distributions of the natural speech training data.  Our analysis method requires typical listening test datasets of audio samples from several different speech synthesis systems all trained on the same natural speech corpus, along with listener-annotated MOS ratings for each system.  The respective speech synthesis training corpus for each listening test dataset is also required as reference data.  First, we extract features of audio samples generated by each of the speech synthesis models, from all layers of a given pretrained model (Whisper \cite{whisper}, XLSR-53 \cite{xlsr53}, and mHuBERT \cite{mhubert}).  Next, we do the same for the speech synthesis training audio (reference data).  We use the Wasserstein distance,  
which is a family of metrics used to compare differences between distributions.  
Following TTSDS \cite{minixhofer2024ttsds}, FID \cite{FID}, and FAD \cite{FAD}, we compute the 2-Wasserstein distance $(W_2)$ between two normally-distributed multi-dimensional distributions $\hat{P}_1$ and $\hat{P}_2$, defined as:  
\begin{equation}
  W_2(\hat{P}_1,\hat{P}_2) = \sqrt{||\mu_1 - \mu_2||^2 + D_B(\Sigma_1,\Sigma_2)}
\end{equation}
where $\mu_1$ and $\mu_2$ are their respective mean vectors, $\Sigma_1$ and $\Sigma_2$ are their covariance matrices, and $D_B$ is the unnormalized Bures metric: 
\begin{equation}
  D_B(\Sigma_1,\Sigma_2) = \mathrm{trace } (\Sigma_1 + \Sigma_1 - 2(\Sigma_2^{1/2}\Sigma_1\Sigma_2^{1/2})^{1/2})
\end{equation}

This distance is computed to measure the distributional difference between each TTS system and the reference data, at each layer.  Finally, at each layer, having obtained $W_2$ for each TTS system, we compute correlations between those system-level $W_2$ distances and the system-averaged MOS ratings to identify which layers have the strongest correlations and are therefore most relevant to listener opinions.  

This layer-wise analysis approach was inspired by an approach \cite{LivescuLWA} in which canonical correlation analysis (CCA) was measured between layer representations extracted from an SSL model and various continuous-valued features such as word embeddings and acoustic features.  While this approach is non-parametric and avoids the need to train classifiers, there is still a minimum data requirement for CCA which is too large for existing MOS datasets.  Therefore, we use our above-described reference-modeling approach using TTS training data as a method for layer-wise analysis that is better-suited to our task and data conditions.

\section{Experiments}

We describe our layer-wise analysis experiments using several different large-scale pretrained multilingual models for speech, as well as several MOS datasets for different languages.

\subsection{Pretrained speech models}

Since the authors of \cite{gui2024adapting} observed that the model chosen can have an effect on the evaluation outcomes, we investigate several pretrained models in this work, but since \cite{tailleur2024correlation} also found that choosing models that are too far from the target domain does not produce reliable correlations with human perceptual ratings, we restrict ourselves to models that were trained specifically using multilingual speech and no other types of audio.

We use two pretrained multilingual speech SSL models, mHuBERT \cite{mhubert}\footnote{github.com/utter-project/mHuBERT-147-scripts} and XLSR-53 \cite{xlsr53}\footnote{github.com/facebookresearch/fairseq/tree/main/examples/wav2vec}.  mHuBERT is based on the HuBERT architecture and is composed of 12 transformer layers trained using a masked prediction training objective on data from 147 languages.  XLSR-53 is based on the wav2vec2 \cite{baevski2020wav2vec2} \texttt{large} architecture with 24 transformer blocks trained using a contrastive learning objective on data from 53 languages.  Both HuBERT and wav2vec2 (including the XLSR-53 variant) have previously been shown to be useful for MOS prediction tasks, with good generalization ability across languages \cite{cooper2022generalization}.

We also investigate the encoder of the Whisper  \cite{whisper}\footnote{github.com/openai/whisper} model for automatic speech recognition (ASR).  The \texttt{large-v3} variant was trained on data from 97 languages with an encoder composed of 32 transformer layers.  Features from the final encoder layer have previously been shown to improve speech intelligibility prediction for noisy and enhanced speech \cite{RyanWhisper}.

\subsection{MOS Datasets}

We gathered several MOS-labeled datasets of synthesized speech for which the corresponding TTS training corpus was also available. 
We emphasize that available datasets that meet all of these criteria, especially non-English ones, are extremely few in number. 
We summarize each dataset below.

\textbf{Blizzard Challenge 2023 French:} (NEB-Fr, AD-Fr) \cite{PERROTIN2025101747} This dataset contains French synthesized samples submitted by participants in the Blizzard Challenge 2023.  This challenge had two tracks with data from two separate speakers: ``NEB'' (51 hours), the main track, and ``AD'' (2 hours), intended as a speaker adaptation track.  Listening tests were conducted separately for each track and therefore we treat each track as a separate dataset.  There are 21 systems including TTS and natural speech for NEB, and 17 systems for AD.  In addition to asking listeners to provide MOS ratings for quality, speaker similarity and intelligibility were also evaluated on different sets of audio samples, and a separate MUSHRA test for quality was conducted as well.  In this work, we only use the MOS ratings for quality.  The Blizzard teams submitted audio samples at different sampling rates, and the listening tests were conducted without changing the sampling rates of the audio samples.

\textbf{Japanese speech synthesis:} (F009-Ja) \cite{xw2018} This dataset contains samples synthesized from different combinations of acoustic models and vocoders which were state-of-the-art in 2018, with 10 systems in total.  The Japanese training data is the ATR-XIMERA \cite{kawai2004ximera} corpus, female speaker ``F009'' (60 hours).  Although both 16~kHz and 48~kHz samples were included in the listening test, we only use the 16~kHz portion.

\textbf{Back to the Future:} (BTTF-En) \cite{bttf} This English listening test re-evaluated some Blizzard Challenge 2013 \cite{blizzard2013} systems alongside modern neural systems, with 8 systems in total.  The TTS training corpus consists of 20 hours of recordings of a female American English professional speaker reading audiobooks in various genres.  While all of our other datasets contain only neural TTS, the BTTF listening test also evaluated unit selection, hybrid, and HMM-based TTS alongside neural systems together in the same listening test.  The sampling rate of the audio samples in this listening test was 16~kHz.  

\textbf{ZMM-TTS:} (ZMM) \cite{zmmtts}  Listening tests in several languages were conducted to evaluate the multi-lingual, multi-speaker speech synthesizer ZMM-TTS.  The English, German, Spanish, and Portuguese datasets are multi-speaker and these languages were included in the training of ZMM-TTS.  Italian and Polish were evaluated as ``unseen'' languages, with synthesized samples in the voice of a single speaker per language in zero-shot and low-resource conditions.  The listening test datasets for the initial four languages contain 10 systems each, and there are 15 systems each for Italian and Polish.  In addition to asking listeners to rate naturalness, the Italian and Polish listening tests asked for ratings of speaker similarity (compared to a reference sample) and intelligibility, at the same time and for the same audio samples.
The training data for ZMM-TTS includes Multilingual LibriSpeech (MLS) \cite{Pratap2020MLSAL}, single-speaker datasets for some of the languages (LJSpeech \cite{ljspeech17} for English; CSS10 \cite{park2019css10} for German and Spanish), and GlobalPhone \cite{schultz2013globalphone}.  Since we did not hold a license for GlobalPhone at the time of writing this paper, we did not include this portion of the ZMM-TTS training data in our reference models.  The sampling rate for audio samples in the listening tests for all languages was 16~kHz.

All listening tests contained held-out natural speech samples in addition to synthesized ones; in each dataset we thus consider natural speech as one of the ``systems.''  All audio data was downsampled to 16~kHz\footnote{While we are aware that downsampling removes salient information at higher frequencies that likely affected listener opinions in the Blizzard Challenge French listening tests, this downsampling was necessary because our pretrained large multilingual speech models only accept 16k~Hz input.  We take this as an opportunity to observe the effect of removing this information on the predictability of listener opinions using these models.}, normalized using the \texttt{sv56demo}\footnote{github.com/openitu/STL/tree/dev/src/sv56} implementation of ITU P.56 \cite{sv56}, and leading and trailing silences were trimmed and padded using SoX\footnote{sourceforge.net/projects/sox/}.  Listener ratings were for either naturalness or overall quality (it has been shown that the choice between these makes little difference in the overall ranking of systems \cite{mospit}).  
A summary of the datasets can be found in Table \ref{tab:datasets}.

\begin{table}[t]
\footnotesize
	\centering
	\caption{Information about MOS datasets and the corresponding available training data used to create reference models.  Number of TTS systems, samples per system, and hours of training (reference) data are shown.}
	\vspace{5pt}
	\centering
	\begin{tabular}{ c c c c}
		\toprule
		Dataset & \# sys & \# samp/sys & Ref hrs \\
  		\midrule
        NEB-Fr & 21 & 42 &  51 \\
        AD-Fr & 17 & 42 & 2 \\
        F009-Ja & 10 & 500 & 60 \\
        BTTF-En & 8 & 16 & 20 \\
        ZMM-En & 10 & 36 & 23 \\
        ZMM-Ge & 10 & 36 & 10 \\
        ZMM-Sp & 10 & 36 & 12 \\
        ZMM-Por & 10 & 36 & 4 \\
        ZMM-It & 15 & 8 & 39 \\
        ZMM-Pol & 15 & 8 & 65 \\
		\bottomrule
	\end{tabular}
	\label{tab:datasets}
\end{table}

\subsection{Layer-wise analysis for quality MOS}

\begin{figure*}[t]
\begin{center}
\includegraphics[width=1.0\textwidth]{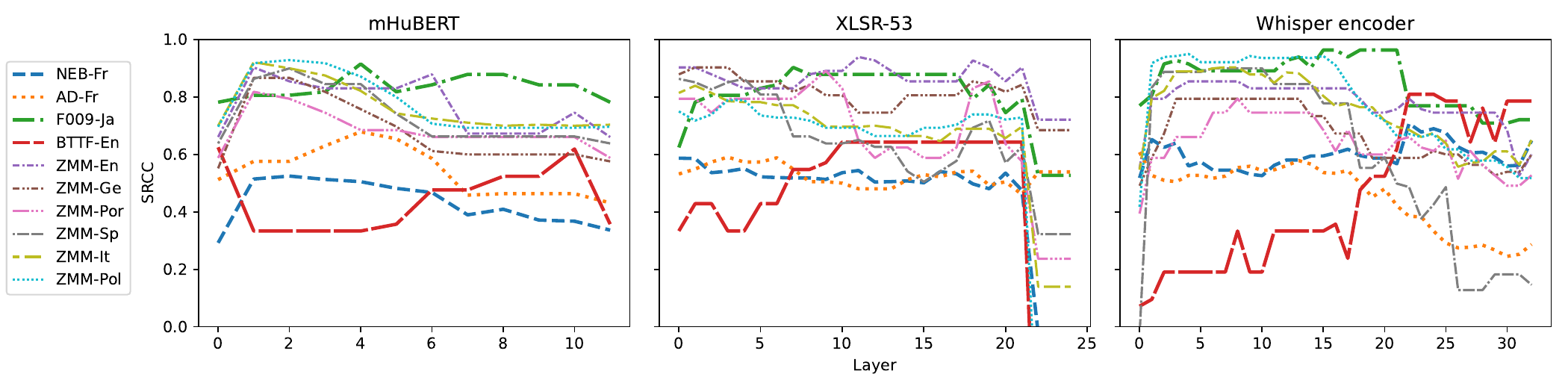}
\end{center}
\vspace{-20pt}
\caption{Layer-wise correlations between MOS and $W_2$ of feature distributions extracted from natural and synthesized speech.}
\vspace{-17pt}
\label{fig:fig1}
\end{figure*}

In our first set of layer-wise analysis experiments, we measure correlations of the $W_2$ distances between synthesized and natural speech with ground-truth system-level MOS ratings for quality or naturalness for three different large pretrained multilingual speech models.  Results are shown in Fig. \ref{fig:fig1}.  Well-matched features are expected to have low $W_2$ distances and negative correlations with MOS, so we show and report {\em negated} correlations for more intuitive readability.  In the XLSR-53 plot, values below 0 have been cut off to avoid compressing the rest of the graph, but the NEB-Fr, ZMM-Pol, and BTTF-En datasets follow the same pattern as the other datasets of decreasing and then leveling off, at values of $-0.04$, $-0.44$, and $-0.74$, respectively. 

Notably, we observe that BTTF-En has very different or even opposite tendencies from the other datasets.  We hypothesize that this is because unit selection and HMM synthesizers have different issues affecting their perceived quality than neural ones.  Given that stronger correlations for this dataset appear in the later layers of the Whisper encoder, which was trained for ASR, we suspect that there may have been more intelligibility issues affecting listeners' naturalness ratings for this dataset, as opposed to the datasets containing only neural systems, which may have fewer intelligibility issues but other issues such as signal quality or prosody.  As for the other datasets containing neural TTS only, we can observe that the $W_2$ distances from early layers have strong and relatively consistent correlations with ground-truth system rankings.  

Both French datasets are more difficult to predict, matching previous findings on these same datasets \cite{vmc2023}.  This is likely due to the loss of salient high-frequency information by downsampling; further investigation on this point is left to future work.

\subsection{Comparison with selected existing models}

We selected several prior MOS prediction approaches to determine their performance on different languages in order to compare them to our reference modeling based approach. We report Spearman rank correlations (SRCC) between the predicted and ground-truth system-level MOS values.

We chose the following pretrained and open-source MOS predictors as representative approaches for comparison:

\textbf{UTMOS:} \cite{saeki22c_interspeech}\footnote{github.com/sarulab-speech/UTMOS22}  A supervised ensemble system of strong learners (SSL models) and weak learners (traditional regression models) trained on English MOS data.  UTMOS has become a popular objective evaluation measure for TTS (e.g., \cite{gongIS24}) and although the authors have since released a newer model, UTMOSv2 \cite{baba2024utmosv2}, our preliminary experiments revealed that the original version had better generalization ability to different languages.

\textbf{NISQA-TTS:} \cite{mittag20_interspeech}\footnote{github.com/gabrielmittag/NISQA}  
A supervised CNN-LSTM architecture trained on MOS datasets in several languages.  To our knowledge, this is the only open-source multilingual-trained MOS predictor, and it has also become a popular choice for objective evaluation (e.g., \cite{10792914}). 

\textbf{VQScore:} \cite{fuVQscore}\footnote{github.com/JasonSWFu/VQscore} A VQ-VAE trained on clean natural English speech with reconstruction error used as the quality measure, chosen as a representative unsupervised approach that has been shown to generalize well to other languages.

\textbf{Entropy:} \cite{aditya} We chose the entropy of SSL token distributions as a representative zero-shot approach using metrics extracted from SSL models (XLSR-53 in this case).  Since entropy is expected to be inversely related to MOS, we report negated correlations for readability.

Prediction results for each of these systems on our selected datasets can be found in Table \ref{tab:predictions}.   High correlations can be observed in several out-of-domain cases; however, none of the predictors have consistent results across all languages.  We also report our best layer results from each layer-wise analysis model, along with which layer(s) it came from.

We can see that many correlations for mHuBERT and XLSR-53 are above $0.8$, and, looking back at Fig. \ref{fig:fig1} at the early layers of mHuBERT and XLSR-53, that that the worst correlations for the neural-only datasets in those early layers are all above 0.5, which are better than the worst correlations from any of the pretrained predictors in Table \ref{tab:predictions}.  This indicates that $W_2$ distances of features extracted from early layers of pretrained models may be a more {\em reliable} predictor of MOS for neural TTS in a previously-unseen language than supervised systems trained for other languages.  As for BTTF-En, the later layers of Whisper produced an SRCC of $0.810$, higher than UTMOS, and without any MOS-specific training or data requirements.

\begin{table*}[t]
\footnotesize
	\centering
	\caption{Left side: SRCC of various MOS predictors.  All datasets are unseen; bold numbers indicate that MOS-labeled data for that language was present in the predictor's training data.  Right side: SRCCs (and best layers) from  layer-wise analysis are shown.}
	\vspace{-5pt}
	\centering
	\begin{tabular}{ c | c c c c | c c c }
		\toprule
		Dataset & UTMOS & NISQA-TTS & VQScore & Entropy & mHuBERT & XLSR-53 & Whisper \\
		\midrule
        NEB-Fr & $0.630$ & $0.345$ & $0.134$ & $0.373$ & $0.525$ (2) & $0.587$ (0) & $0.708$ (22) \\
        AD-Fr & $0.287$ & $0.706$ & $0.213$ & $0.475$ & $0.679$ (4) & $0.591$ (3) & $0.608$ (0) \\
        F009-Ja & $0.879$ & $0.842$ & $0.467$ & $-0.467$ & $0.915$ (4) & $0.903$ (7) & $0.964$ (15-18,19-21) \\
        BTTF-En & $\mathbf{0.786}$ & $\mathbf{0.714}$ & $\mathbf{-0.071}$ & $0.286$ & $0.625$ (0) & $0.643$ (10-21) & $0.810$ (22-24) \\
        ZMM-En & $\mathbf{0.806}$ & $\mathbf{0.964}$ & $\mathbf{0.685}$ & $0.600$ & $0.903$ (1) & $0.939$ (11) & $0.855$ (4-8) \\
        ZMM-Ge & $0.830$ & $\mathbf{0.830}$ & $0.491$ & $0.224$ & $0.867$ (1,2) & $0.903$ (1-3) & $0.794$ (5-13) \\
        ZMM-Sp & $0.681$ & $0.778$ & $0.365$ & $0.024$ & $0.900$ (2) & $ $0.863$ (0,4) $ & $0.900$ (6-10) \\
        ZMM-Por & $0.891$ & $0.903$ & $0.564$ & $0.297$ & $0.818$ (1) & $0.891$ (9) & $0.794$ (8) \\
        ZMM-It & $0.871$ & $0.346$ & $0.364$ & $0.036$ & $0.921$ (1) & $0.839$ (1) & $0.904$ (7,8) \\
        ZMM-Pol & $0.843$ & $0.336$ & $0.757$ & $0.332$ & $0.929$ (2) & $0.789$ (3,4) & $0.943$ (3,9,15) \\
		\bottomrule
	\end{tabular}
 \vspace{-15pt}
	\label{tab:predictions}
\end{table*}

\subsection{Speaker similarity and intelligibility}

\begin{figure}[t]
\begin{center}
\includegraphics[width=1.0\columnwidth]{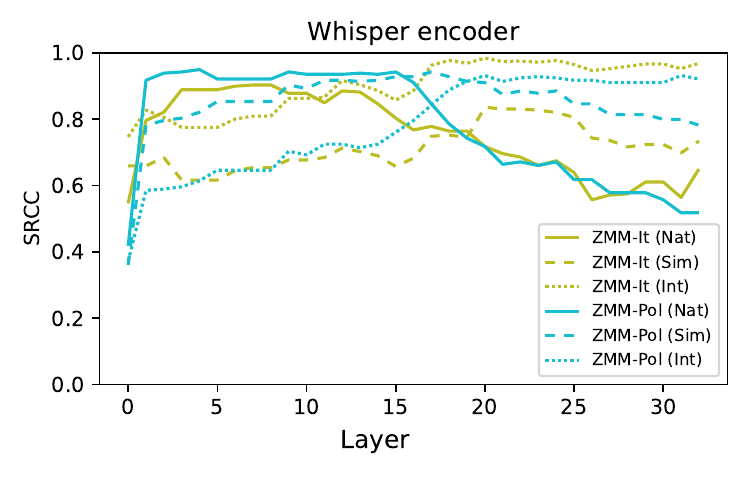}
\end{center}
\vspace{-25pt}
\caption{Layer-wise correlations for naturalness, speaker similarity, and intelligibility.}
\vspace{-10pt}
\label{fig:fig2}
\end{figure}

The ZMM Italian and Polish listening tests also asked listeners for ratings of speaker similarity and intelligibility, so we have ratings for those factors in addition to naturalness for the same set of audio samples.  We conducted the same layer-wise analysis as before and show the correlations with those different sets of ratings in Fig. \ref{fig:fig2}, focusing on the Whisper encoder because we expect it to show a clear pattern for intelligibility.  We can indeed see that correlations with intelligibility increase towards later layers consistently for both languages.  Speaker similarity shows a peak towards the middle but with a less consistent pattern.  Different pretrained models such as speaker encoders are probably better-suited for the speaker similarity task, but we leave layer-wise analysis of speaker models for future work.

\subsection{Importance of matched reference data}

It has been shown in earlier work on FAD \cite{gui2024adapting} that the reference data should be properly chosen in order to get meaningful correlations with subjective ratings.  In order to investigate what types of reference data might be suitable to use instead of the original TTS training data, we conducted a set of experiments using various reference datasets with different mismatched conditions.

\begin{figure}[t]
\begin{center}
\includegraphics[width=1.0\columnwidth]{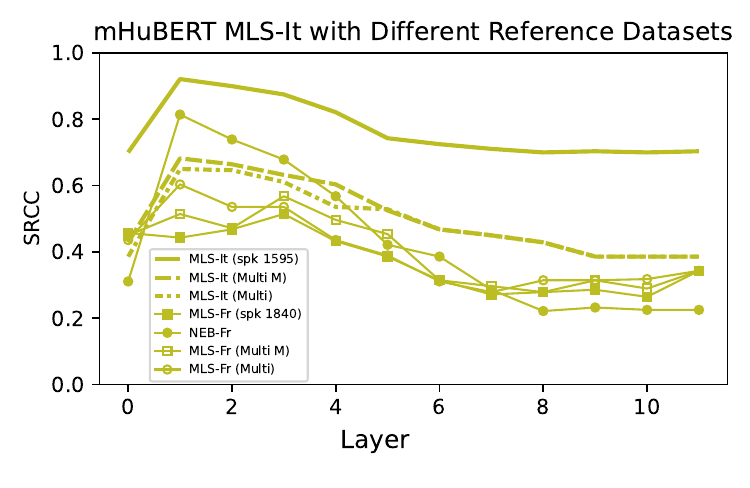}
\end{center}
\vspace{-25pt}
\caption{Effects of choice of reference modeling data.}
\vspace{-13pt}
\label{fig:fig3}
\end{figure}

We picked ZMM Italian as a representative single-speaker dataset and mHuBERT as a representative SSL model with a clear layer-wise pattern for quality MOS.  The TTS systems in the ZMM-It dataset were trained on the MLS Italian male speaker \#1595, the speaker with the most data, 39 hours.  We looked at using datasets of a similar size but that were mismatched in various ways to observe how various types of mismatch affect the correlations: 39 hours each of male-only Italian multi-speaker data; mixed-gender Italian multi-speaker data; the French NEB training data to match the single-speaker setting and audiobook domain but not speaker ID, language, or gender; MLS French male speaker \#1840; and both male-only and mixed-gender French multi-speaker data.  Results are shown in Fig. \ref{fig:fig3}.  It can be observed that the matched condition of using the training corpus as reference data clearly produces the best correlations with MOS across all layers.  Interestingly, the language- and gender-mismatched NEB data has the second-best correlation in the region of the early layers but degrades to have the worst correlation at the later layers.  As we would expect, the multi-speaker Italian data tends to have slightly better correlations than the multi-speaker French data across most layers.  Using Italian matched-gender multi-speaker reference data only appears to give a small benefit over using mixed-gender reference data.  We can therefore recommend using the matched TTS training data as reference data whenever possible.

\section{Discussion and conclusions}

We conducted a layer-wise analysis of pretrained multilingual speech models for correlations with various aspects of listener opinions using a reference modeling based approach.  Computing $W_2$ distances between distributions of synthesized speech and natural speech, followed by the correlations between those distances and MOS, we found that early layers correlate well with opinions of overall naturalness and this measure can therefore be used as a proxy for predicted MOS without any additional MOS-specific training.  We note that care must be taken when using this approach for evaluating non-neural TTS systems, as datasets including these showed different tendencies, with later layers tending to produce better correlations.  This approach is probably also not suitable for evaluating synthesized speech at sampling rates higher than 16~kHz, since our chosen pretrained models only support that sampling rate and information salient to listener opinions does appear to have been lost by downsampling in the case of the French data.  
Important future work will be to determine principled ways to choose the best layer (or combination of layers) in real predictive scenarios in order to use the knowledge obtained in this study to develop unsupervised multilingual quality predictors.  

We observed that using matched reference data worked better than reference data containing various types of mismatches.  However, a main limitation of our approach is that matched reference data may not always be available for the target synthesis, e.g. for language or style transfer TTS applications where ground-truth data in a particular combination of language, target speaker, and speaking style may not exist. We also note that all of the languages of our datasets were present in the pretraining data for all three large speech models.  It will be important future work to investigate layer-wise predictive ability for low-resource languages which have been completely unseen by these models.

\section*{Acknowledgment}

We thank Prof. Junichi Yamagishi for the use of the Japanese and ZMM-TTS listening test data, as well as the organizers of the Blizzard Challenge for making their data available.  We also thank Marcely Zanon Boito, Gabriel Mittag, Szu-Wei Fu, and Soumi Maiti for the helpful discussions about the use of their shared code and models.

\bibliographystyle{IEEEtran}
\bibliography{mybib}

\end{document}